\documentclass[prl,superscriptaddress,twocolumn,showpacs]{revtex4}
\usepackage{graphicx}
\usepackage[cp1250]{inputenc}            

\begin{document}

\title{Spin filtering in a hybrid ferromagnetic-semiconductor microstructure}

\author{J. Wróbel}
\author{T. Dietl}

\affiliation{Institute of Physics, Polish Academy of Sciences, al.
Lotników 32/46, 02-668 Warszawa, Poland} \affiliation{ERATO
Semiconductor Spintronics Project, Japan Science and Technology
Agency, al.~Lotnik\'ow 32/46, 02-668 Warszawa, Poland}

\author{A. Łusakowski}

\affiliation{Institute of Physics, Polish Academy of Sciences, al.
Lotników 32/46, 02-668 Warszawa, Poland}

\author{G. Grabecki}

\affiliation{Institute of Physics, Polish Academy of Sciences, al.
Lotników 32/46, 02-668 Warszawa, Poland} \affiliation{ERATO
Semiconductor Spintronics Project, Japan Science and Technology
Agency, al.~Lotnik\'ow 32/46, 02-668 Warszawa, Poland}

\author{K. Fronc}

\affiliation{Institute of Physics, Polish Academy of Sciences, al.
Lotników 32/46, 02-668 Warszawa, Poland}

\author{R. Hey}
\author{K. H. Ploog}

\affiliation{Paul Drude Institute of Solid State Electronics,
Hausvogteiplatz 5-7, D-10117 Berlin, Germany}

\author{H. Shtrikman}
\affiliation{Center for Submicron Research, Weizmann Institute of
Science, Rehovot 76100, Israel}

\begin{abstract}
We fabricated a hybrid structure in which cobalt and
permalloy micromagnets produce a local in-plane spin-dependent
potential barrier for high-mobility electrons
at the GaAs/AlGaAs interface. Spin effects are observed in
ballistic transport in the tens' millitesla range of the external
field, and are attributed to switching between Zeeman
and Stern-Gerlach modes -- the former dominating at low electron
densities.
\end{abstract}


\maketitle

Long spin coherence times in semiconductors have triggered
considerable efforts towards developing devices, in which
functionalities would involve spin degrees of freedom
\cite{Awsch02}. An important building block of such devices is a
spin-filter, which could serve for either generating or detecting
spin-polarized currents and, indeed, spin filtering capabilities
of quantum point contacts \cite{Potok02,Grabec02} and quantum dots
\cite{Folk03,Ciorga02,Hanson03} have recently been demonstrated.
In those devices, a spin dependent barrier occurs as a result of
the Zeeman spin-splitting generated by a strong uniform
\emph{external} magnetic field.  Also the Stern-Gerlach (S-G)
effect has been theoretically considered as a possible spin-filter
in spin-logic processors \cite{Barn00}. There are, however,
fundamental arguments against the occurrence of the S-G effect for
beams of electrons \cite{Mott29}, a problem that attracts
persistent attention \cite{Bat97,Garr99,Gall01}. At the same time,
the progress in fabrication of hybrid ferromagnet-semiconductor
microstructures \cite{John97,Matsu02} has made it possible to
address various aspects of electron transport in the presence of
an inhomogeneous magnetic field
\cite{Nog00,Law01,Badal01,Wro01,Fab02,Grabec04}. For example, the
present authors have proposed a way to achieve spin separation by
using a Stern-Gerlach apparatus for the conduction electrons
residing in a quantum well and exposed to a gradient of the
in-plane magnetic field \cite{Wro01}.

In this Letter, we report on the effect of a {\em local} in-plane
magnetic field on ballistic currents in a quantum wire patterned
of GaAs/GaAlAs heterostructure. The results are obtained for a
ferromagnet-semiconductor hybrid device which is highly optimized
in order to toggle between Zeeman-like (uniform field) and
Stern-Gerlach-like (field gradient) internal spin barriers. By
comparing our findings to results of conductance computations by
the recursive Green-function method, we find out that the Zeeman
effect dominates, particularly at low carrier densities. Owing to
spin filtering and detecting capabilities that occur in a weak
external magnetic field, our microstructure thus emerges as a
perspective component of spintronic devices.

Figure 1(a) presents a micrograph of our device, whose design
results from an elaborated optimization process \cite{Wro01}, and
whose fabrication involves five electron beam lithography
levels, two wet etching steps, and deposition by low-power
magnetosputtering and lift off of four different metals. A
two-dimensional electron gas (2DEG) resides 95~nm below the top
surface of a quantum wire of the geometrical width smoothly
increasing from 0.7 to 1.4~$\mu$m, chemically etched from a
modulation Si-doped (001) GaAs/(Al,Ga)As heterostructure, which
was grown in the Drude Institute in Berlin. The wire is patterned
along the [110] crystal axis, for which the direction of a
fictitious magnetic field brought about by spin-orbit effects will
be parallel to the field gradient and, therefore, will weakly
affect spin dynamics \cite{Lusak03}. The electron mobility prior
to nanofabrication is $\mu = 1.76\times 10^6$ cm$^2$/Vs in the
dark. Hence, with the electron density $n = 2.3\times 10^{11}$
cm$^{-2}$, the mean free path is of the order of the channel
length. A local magnetic field is produced by NiFe (permalloy, Py)
and cobalt (Co) films. The micromagnets of dimensions $40\times
7\times 0.1$~$\mu$m$^3$ reside in $0.15 \pm 0.05$~$\mu$m deep
groves on the two sides of the wire, so that the 2DEG is
approximately at the center of the field. To prevent stripe
oxidation and to avoid accumulation of electrostatic charges, both
micromagnets are covered with a thin (20 nm) protecting AuPd layer
and connected to separate contact pads. Additional narrow groves
patterned on the wire entrance and exit define electron emitter
and two counters, respectively. Annealed films of AuGe constitute
ohmic contacts between the 2DEG and current leads. The
micromagnets are also used as side gates which, together with
illumination by infrared radiation, serve for controlling the
number of occupied 1D subbands.

\begin{figure}
\includegraphics[scale=0.31,angle=-90]{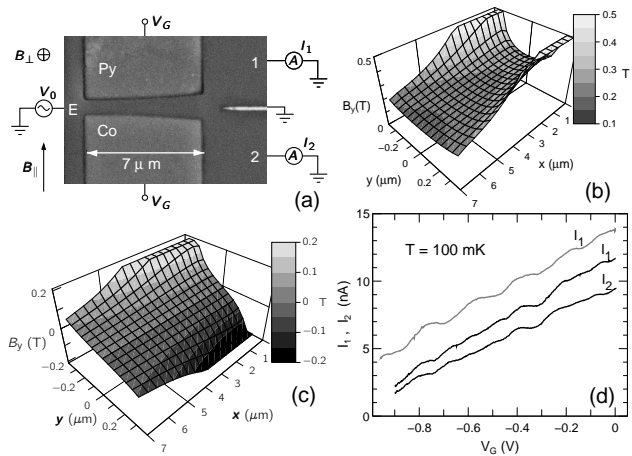}
\caption{(a) Scanning electron micrograph of the spin-filter
device. Fixed AC voltage $V_0$ is applied between emitter (E) and
"counters" (1), (2); DC gate voltage $V_G$ is applied to both
magnets which act as side gates. In-plane magnetizing field
($B_\parallel$) and perpendicular field ($B_\perp$) are oriented
as shown. (b) The in-plane magnetic field $B_y$ (wider part of the
channel is in front) calculated for half-plane, 0.1~$\mu$m thick
magnetic films separated by a position dependent gap $W(x)$ and
magnetized in the same directions (saturation magnetization as for
Co). (c)$B_y$ calculated for antiparallel directions of
micromagnet magnetizations. (d) Counter currents $I_1$ and $I_2$
as a function of the gate voltage at $V_0=100$~$\mu$V; upper curve
(shown in gray) was collected during a different thermal cycle and
after longer infra-red illumination.} \label{fig1}
\end{figure}

We have applied Hall magnetometry in order to visualize directly
the magnetizing process of the two micromagnets in question. Hall
microbridges, patterned of GaAs/(Al,Ga)As:Si heterostructures
grown in the Weizmann Institute in Rehovot, contain a 2DEG at 47.5
nm below the top surface on which  Py and Co micromagnets,
analogous to those in the spin-filter device, are deposited.
Figure 2 presents Hall resistance as a function of the in-plane
magnetic field for three bridges which contain either single
micromagnets or a pair of them. Step-like changes of the Hall
resistance are caused by a consecutive reversal of magnetic
domains.  According to Fig. 2, the Co and Py micromagnets have
differing coercive fields but similar saturation magnetizations
$M_s$.  Thus, in the spin filter device, we can compare the
electric currents through the counters in the presence of the
virtually uniform magnetic field (parallel magnetization
directions) to the case when a strong field gradient is present.
The spatial distribution of the magnetic field, evaluated under
assumption that the values of $M_s$ correspond to that of Co,
$\mu_oM_s = 0.179$~T, are presented in Figs. 1(b) and 1(c). We see
that depending on the relative magnetization directions the
electrons will experience the local magnetic field $B$ up to 0.3 T
[Fig. 1(b)] or the local field gradient up to $10^6$ T/m [Fig.
1(c)].

\begin{figure}
\includegraphics[scale=0.31,angle=0]{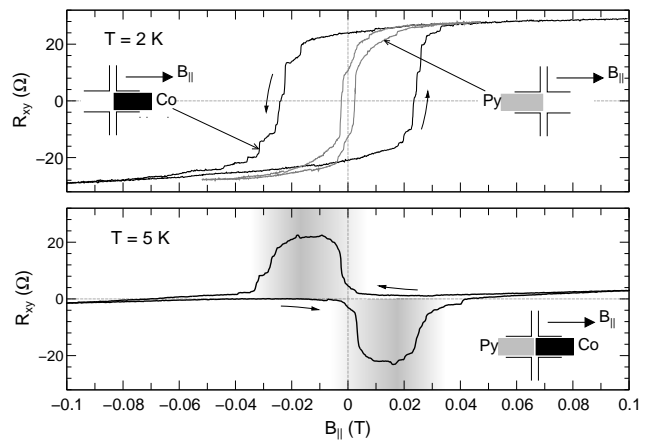}
\caption{ Hall resistance as a function of in plane magnetic field
measured for devices with single Co magnet, single Py magnet (top)
and with two magnets separated by $1~\mu$m gap (bottom). The
arrows indicate directions of the magnetic field sweep. Shaded
bands in lower panel denote the magnetic field range where the
magnetizations of Co and Py micromagnets are antiparallel.}
\label{fig2}
\end{figure}

\begin{figure}
\includegraphics[scale=0.43,angle=0]{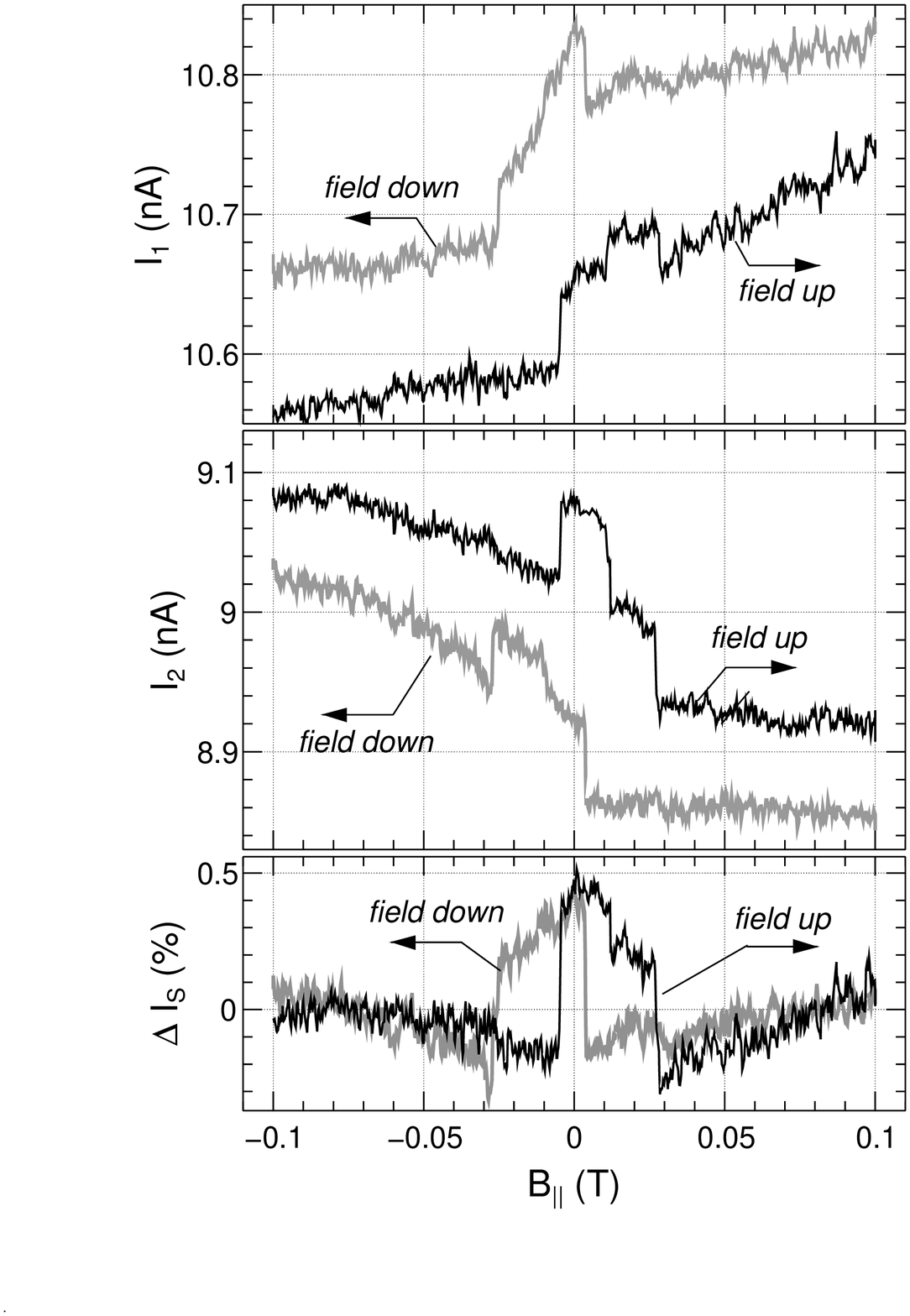}
\caption{ Counter currents $I_1$ and $I_2$ at bias $V_0 =
100$~$\mu$V and at 100 mK as a function of the in-plane magnetic
field at zero gate voltage. The arrows indicate directions of the
magnetic field sweep. The data for field down sweep are shifted up
(by 0.05 nA) for clarity. The relative changes of symmetric
component of the signal $I_S=(I_1+I_2)/2$, which eliminates the
Hall effect, are shown in the bottom panel. $\Delta I_S$ is
defined as $(I_S-I^{++}_S)/I^{++}_S$, where
$I^{++}_S=I_S(B_\parallel=0.1$~T$)$.} \label{fig3}
\end{figure}

Our electron transport measurements for the spin filter device are
carried out in a dilution refrigerator at 100 mK employing a
standard low-frequency lock-in technique. According to results
presented in Fig. 1 (d), conductance plateaux are clearly
resolved. Their heights imply that the total transmission
coefficient is about 0.7, a value consistent with the presence of
the reflecting barrier separating the two counters. Since during
these measurements micromagnets were not magnetized, a visible
difference in counter currents provides information about the
degree of structure symmetry. What should we expect when the
spin-dependent potential barriers are switched on? Classically,
the presence of the S-G effect should manifest itself by a
gradient-induced symmetric enhancement of the current through the
both counters at given emitter-counter and gate voltages. As shown
in Fig.~\ref{fig3}, we detect a current increase in both counters
when a field gradient is produced by an appropriate cycle of the
external magnetic field. The range of magnetic fields where the
enhancement is observed corresponds to the shaded bands in
Fig.~\ref{fig2}(b). Furthermore, the magnitude of the stray field
produced by Py is seen in Fig.~\ref{fig2}(a) to diminish almost
twofold prior to a change in the direction of the external
magnetic field. This effect, associated with the formation of
closure domains in soft magnets, explains why the current changes
appear before the field reversal. The current enhancement in
question is superimposed on a slowly varying background, which
exhibits antisymmetric behavior  for the two counters. We assign
its presence to a residual effect of the Lorentz force (Hall
effect) associated with a possible misalignment of the
micromagnets in respect to the 2DEG plane.

\begin{figure}
\includegraphics[scale=0.33,angle=0]{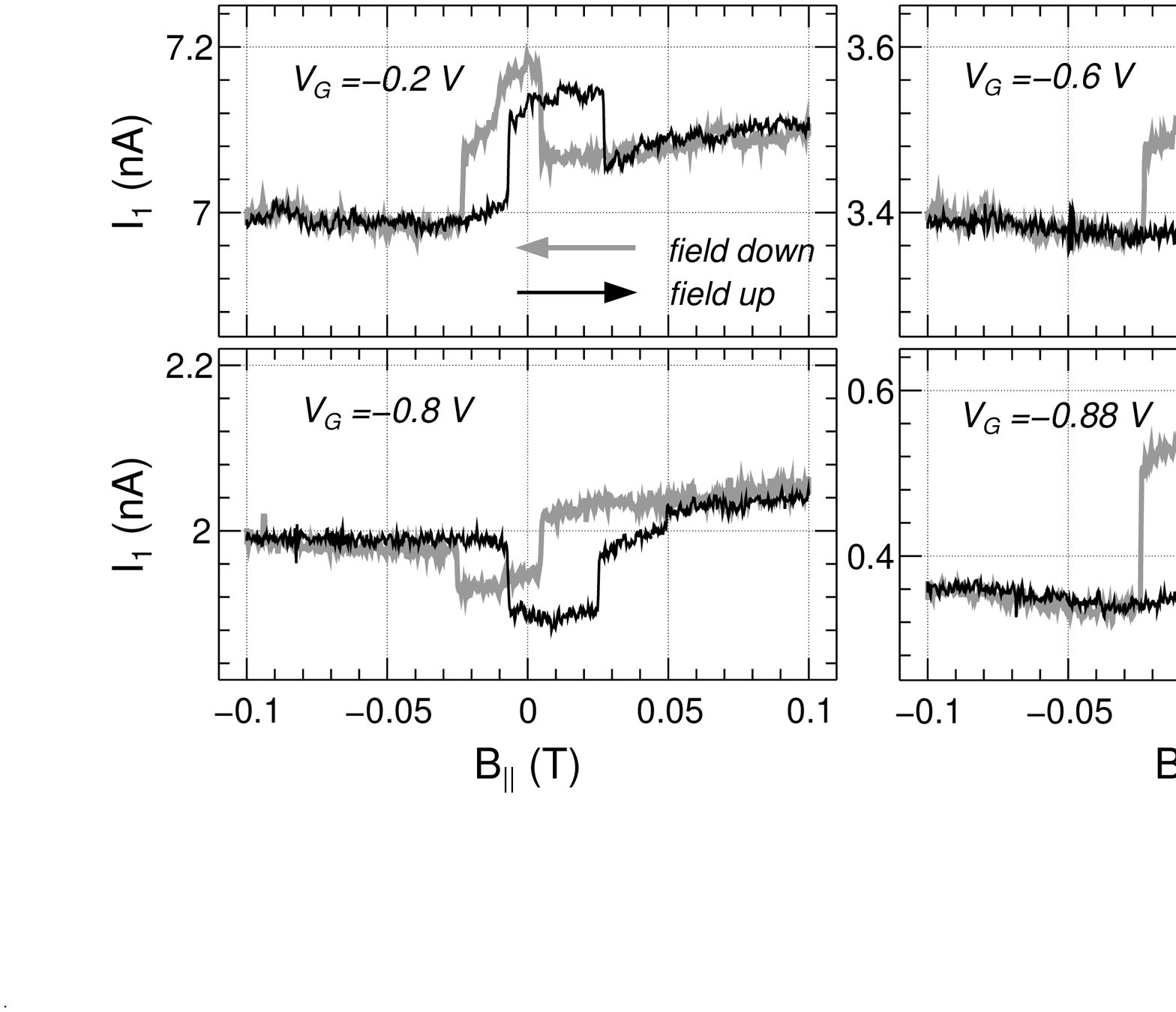}
\caption{The counter current $I_1$ as a function of the in-plane
magnetic field for various gate voltages.} \label{fig4}
\end{figure}

We checked that results presented in Fig.~\ref{fig3} are
unaltered by increasing the temperature up to 200 mK and
independent on the magnetic field sweep rate. The relative change
$\Delta I$ of counter current depends, however, on the gate
voltage $V_G$. Figure ~\ref{fig4} shows $I_1$ vs $B_\parallel$ for
several values of $V_G$ ($I_2$ behaves in the same manner).
$\Delta I/I$ increases form $0.5$~\% at zero gate voltage to
$50$~\% close to the threshold. Furthermore, for $V_G$ about
$-0.8$~V $\Delta I$ is negative.

\begin{figure}
\includegraphics[scale=0.43,angle=0]{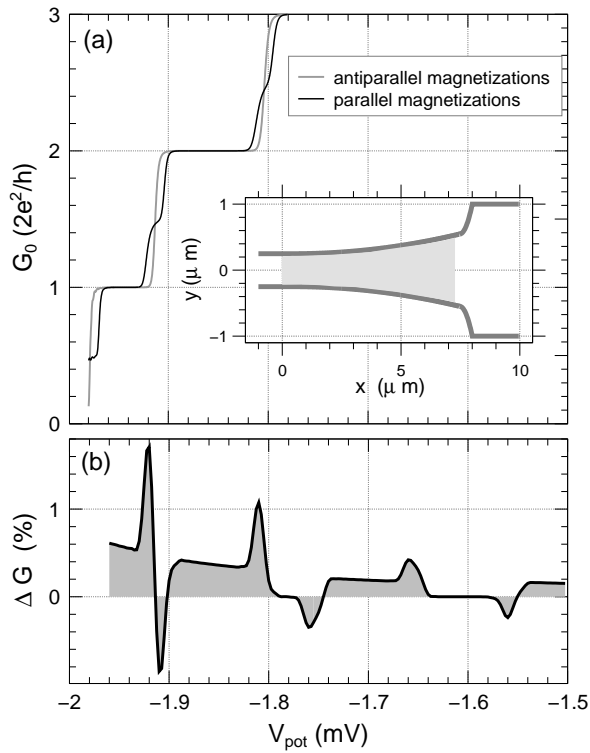}
\caption{(a) Quantized zero bias conductance $G_0$ calculated for
parallel (black) and antiparallel (grey) magnetizations . The
model potential $V(x,y)$ is shown in the inset. Thick dark gray
line denotes the hard wall potential. Within the light grey
region, an additional electrostatic potential $V_{pot}$ is
adiabatically introduced to simulate the gate potential.
Conductance is shown as a function of $V_{pot}$ for electron
energy $E=2$~meV and minimal channel width $W_0=0.5$~$\mu$m. (b)
Relative changes of the conductance $G$ for the , defined as
$\Delta G=(G^{+-}-G^{++})/\langle G \rangle$, where $G^{+-}$
corresponds to the Stern-Gerlach and $G^{++}$ to the spin-filter
configurations respectively; $\langle G \rangle$ is the average
conductance for both configurations.} \label{fig5}
\end{figure}

We evaluate the expected magnitude of $\Delta I/I$ within the
model of quantum ballistic transport, which we developed
previously \cite{Lusak03} by employing the recursive Green
function method. We note that the key feature of our experimental
configuration is a dramatic reduction of the influence of the
Lorentz force by electron confinement. In particular, the effect
of the in-plane magnetic field $B_{x,y}$ is much reduced by the
interfacial electric field and the corresponding quantization of
electron motion in the $z$ direction. Furthermore, since a
residual field $B_z$ brought about by misalignment of the magnet
centers tends to vanish in the branching region, its influence on
electron dynamics will be small \cite{Been89}, in agreement with
with a weak asymmetry of data in Figs. 3 and 4. Under these
assumptions, electron dynamics is governed by the potential
$V(x,y)$ determined by the device geometry, taken in the form
shown in the inset to Fig.~\ref{fig5} as well as by the Pauli
term, $g^*sB$, where $B=(0,B_y(x,y),0)$ with $B_y(x,y)$ displayed
in Figs.~\ref{fig1}(b,c) for both magnetization configurations.
Because of low density of electrons in the quantum wire, we expect
a considerable enhancement of the electron Land\'e factor. The
interaction induced renormalization of the $g$-factor has been
already observed experimentally for the gated low-density 2D
electron gas \cite{Tutuc02,Zhu03}. While the role of many body
effects in confined systems is under an active debate presently,
we take their existence into account by allowing for an
enhancement of the Landé factor to the value $|g^*| = 2.0$.

The zero bias conductance $G_0$ of our model device is shown in
Fig.~\ref{fig5}(a) as a function of the electrostatic potential
barrier height. Black and grey lines correspond to the Zeeman-like
and Stern-Gerlach-like spin dependent barriers, respectively. As
expected, when both micromagnets are polarized in the same
direction an additional spin-resolved narrow \emph{plateaux} shows
up. Reversing the magnetization of Py pole while leaving the Co
pole unaffected corresponds to the transition from the black to
the grey curve. As a result, at the transition region between
plateaux $\Delta G_0$ is either positive or negative. At the
quantized \emph{plateaux}, the conductance does not depend on the
type of the spin barrier, and the spatial distribution of total
current density is only slightly modified by the presence of the
field gradient. Under these conditions, however, the electric
current at opposite edges of the device is strongly "left" or
"right" spin polarized, up to $50$~\% for $G=1$. This indicates
that the S-G effect is present under our experimental conditions
though it contributes weakly to the current enhancement visible in
Figs. 3 and 4.

It is clear that $\Delta G$ should be averaged over a non-zero
energy window corresponding to the applied emitter/counter
voltage. Quite remarkably, the non-zero bias ($V_0=100$~$\mu$V),
which is $\sim 3$ times larger than the expected spin splitting
($30$~$\mu$eV for $g^*=2$), does not smear out the changes of the
conductance associated with the presence of the Zeeman barrier.
Actually, it extends the regions of conductance changes towards
the quantized \emph{plateaux}. We defined the the observed
conductance as $G=\int_{\mu_1}^{\mu_2}G_0(E)dE$, where $E$ is the
electron energy and $\mu_2-\mu_1=eV_0$. We see that the computed
magnitude of the effect compares favorably with the experimental
findings. Except for the pinch-ff region, where an additional
enhancement of spin-splitting is possible, our model describes the
magnitude of the effect and explains why the sign of the effect
can be negative for some values of the gate voltage.

In conclusion, the experimental and theoretical study presented
here demonstrates that semiconductor nanostructures of the kind
proposed in this work can serve to generate and detect spin
polarized currents in the absence of an external magnetic field.
Moreover, according to our results, the degree and direction of
spin polarization at low electron densities can easily be
manipulated by gate voltage or a weak external magnetic field.
While the results of our computations suggest that the spin
separation and thus Stern-Gerlach effect occurs under our
experimental conditions, its direct experimental observation would
require incorporation of spatially resolved spin detection.

We thank H. Ohno and D. Weiss for valuable discussions. This work
in Poland was partly supported by FENIKS (G5RD-CT-2001-00535) and CELDIS
(ICA1-CT-2000-70018) EC projects as well as by KBN grant
(PBZ-044/P03/2001).

\end{document}